\documentclass[conference]{IEEEtran}
\usepackage{cite}
\usepackage{amsmath,amssymb,amsfonts}
\usepackage[utf8]{inputenc}
\usepackage{indentfirst}
\usepackage{algorithmic}
\usepackage{nomencl}
\makenomenclature
\usepackage{graphicx}% http://ctan.org/pkg/graphicx
\usepackage{subfig}
\usepackage{textcomp}
\usepackage{xcolor}
\def\BibTeX{{\rm B\kern-.05em{\sc i\kern-.025em b}\kern-.08em
    T\kern-.1667em\lower.7ex\hbox{E}\kern-.125emX}}
\begin{document}

\title{Privacy Preserving Techniques Applied to CPNI Data: Analysis and Recommendations}

\makeatletter
\newcommand{\linebreakand}{%
  \end{@IEEEauthorhalign}
  \hfill\mbox{}\par
  \mbox{}\hfill\begin{@IEEEauthorhalign}
}
\makeatother

\author{
Jeffrey Murray Jr, Afra Mashhadi, Ph.D, Brent Lagesse, Ph.D, Michael Stiber, Ph.D \\
Computing and Software Systems \\
University of Washington Bothell, WA \\
jeffmur@uw.edu, mashhadi@uw.edu, lagesse@uw.edu, stiber@uw.edu
}
% }
% \IEEEauthorblockN{Jeffrey Murray Jr}
% \IEEEauthorblockA{Graduate Research Assistant \\
% Cybersecurity Engineering \\
% University of Washington \\
% Bothell, WA \\
% jeffmur@uw.edu}
% \and
% \IEEEauthorblockN{Afra Mashhadi, Ph.D.}
% \IEEEauthorblockA{Assistant Professor \\
% Computing \& Software Systems\\
% University of Washington\\
% Bothell, WA \\
% mashhadi@uw.edu}
% \and
% \IEEEauthorblockN{Brent Lagesse, Ph.D.}
% \IEEEauthorblockA{Associate Professor, \\
% Computing \& Software Systems\\
% University of Washington \\
% Bothell, WA \\
% lagesse@uw.edu}
% \linebreakand %
% \IEEEauthorblockN{Michael Stiber, Ph.D.}
% \IEEEauthorblockA{
% Professor; Associate Dean, \\ 
% Research and Graduate Studies \\
% University of Washington \\
% Bothell, WA \\
% stiber@uw.edu}
% }

\maketitle
\begin{abstract}
With mobile phone penetration rates reaching 90\%, Consumer Proprietary Network Information (CPNI) can offer extremely valuable information to different sectors, including policy makers. Indeed, as part of CPNI, Call Detail Records has been successfully used to provide real-time traffic information, to improve our understanding of the dynamics of people’s mobility and so to allow prevention and measures in fighting infectious diseases, and to offer population statistics. While there is no doubt of the usefulness of CPNI data, privacy concerns regarding sharing individuals’ data have prevented it from being used to its full potential. Traditional de-anonymization measures, such as pseudonymization and standard de-identification, have been shown to be insufficient to protect privacy. This has been specifically shown on mobile phone datasets. As an example, researchers have shown that with only four data points of approximate place and time information of a user,  95\% of users could be re-identified in a dataset of 1.5 million mobile phone users. In this landscape paper we discuss the state-of-the-art anonymization techniques and their shortcomings.  
\end{abstract}
\section{Introduction}
Telecom operators collect enormous amounts of data that is often referred to as Customer Proprietary Network Information (CPNI). One of the main dimensions of such data is Call Data Records which includes fine grain information regarding the location (latitude and longitude) of the caller, and the callee, along with their associated cell towers and the time of the call. This data is not only essential to the billing operations of the telecom industry but is also extremely valuable for third party agencies and researchers. For example, a common need when analyzing real-world data-sets is determining which instances stand out as being dissimilar to all others. Such instances are known as anomalies, and given a high prevalence of normal instances of data, anomalies can be identified using machine learning approaches. In the context of spatial temporal data such as those of Call Data Records, anomalies could correspond to the high load on a specific cell tower at a given time  caused by a natural phenomena such as  events, crowd flow/traffic jam \cite{oliver_mobile_2015, smith_ubiquitous_nodate}, or by adversarial attacks.  Other use cases of such data include detecting similarities and patterns. For example, CDR data has been shown to improve our understanding of the dynamics of citizens’ mobility and communities \cite{hong_towards_2018, oliver_mobile_2020}. \par
While there is no doubt of the usefulness of this data, privacy concerns regarding sharing individuals’ data have prevented it from being used to its full potential. Indeed, data sharing has been an ongoing challenge for telecom operators as even anonymized Call Data Records can reveal sensitive information regarding the location and mobility of an individual.  One example of privacy concerns associated with spatial-temporal data is unicity, where researchers have shown that with only four data points of approximate place and time information of anonymized mobility traces, 95\ of users could still be re-identified in a dataset of 1.5 million mobile phone users \cite{de_montjoye_unique_2013}. \par
To address the privacy concerns in data sharing, prior works either obfuscate the data through a transformation process which often includes coarsening the spatial data to correspond to the Voronoi translation of the antenna as opposed to the exact latitude and longitude. Alternatively, telecom operators have released their data as a pre-aggregated indicator. These indicators can be computed at individual level, corresponding to the number of calls, or aggregated across individuals to capture for example the number of users per tower over time, long or short-term mobility matrices, and matrices of inter-towers communications. \par
In general, these types of transformations affect the quality and quantity of the data and therefore the statistical inferences that can be made and potentially prevent important research questions to be explored. In this paper, we survey the existing methods for anonymization for telecom data. We show that while various technical solutions exist, each solution has its own limitations and applies to specific applications. \par
The rest of this landscape paper is organized as follows: In Section II we will address a single dimension of CPNI known as CDR and the common trade off of published datasets; Section III, we will discuss current state-of-the-art techniques to anonymize sensitive data elements within call data records, demonstrate the effectiveness and pitfalls of these mechanisms and provide relevant examples. In Section IV, we will provide typical example of data release policies that have been successful, and finally Section V will assert our recommendations for the Carrier to consider prior to a data publication.
\section{Background}
\subsection{Private Data}
The EU General Data Protection Regulation (GDPR) defines personal data as “any information relating to an identified or identifiable natural person (‘data subject’)” and applies to any individual within a dataset that includes ‘direct’ and ‘indirect’ identification. This means “an identification number or to one or more factors specific to his physical, physiological, mental, economic, cultural or social identity of that natural person” \cite{european_parliament_and_council_of_european_union_art_2016}. Moreover, the California Consumer Privacy Act (CCPA) gives the users more specific outlines and public knowledge on the usage of personal information that is collected by businesses \cite{noauthor_california_2018}. Information collected by telecommunication service providers frequently hand off service for customers internationally and optimal delivery within the United States. Calls, texts, or use of data is sent out in a network packet. The type of data considered in Call Data Records is within the following jurisdiction of the CCPA: 
\begin{enumerate}
\item Names, Geographical subdivisions, Dates
\item Phone and Fax Numbers
\item Electronic mail addresses
\item Device identifiers and serial numbers
\item Internet Protocol (IP) address numbers
\item Geolocation
\end{enumerate}

In this paper, two categories of the CCPA will be considered: Identifiers and Geo-location.
\textbf{Identifiers} include explicit and quasi-identifiers where explicit identifiers are attributes of a data subject that directly exposes their identity such as: Name, Address, Social Security, and Phone Number, while quasi-identifiers are shared attributes by many individuals such as Race, Gender, Socioeconomic, and Geolocation. \textbf{Geo-location} includes IP addresses, MAC addresses, fine-grained and coarse-grained mobility records, longitude and latitude coordinates, and relative location. 

\subsection{Spatio-Temporal Data Types}
There is a variety of ST data types that the physical world contains, however they differ in the way space and time are used in the process of data collection and representation which lead to different categories of Spatio-Temporal Data Mining (STDM) problem formations \cite{atluri_spatio-temporal_2018}. Within CDR, it is event driven, while the compiled dataset can extract trajectory feature for each user. \par
% see https://www.overleaf.com/learn/latex/Inserting_Images
\begin{figure}[th]
\caption{Illustration of even data and trajectory data \cite{wang_deep_2019}}
\includegraphics[width=8.75cm]{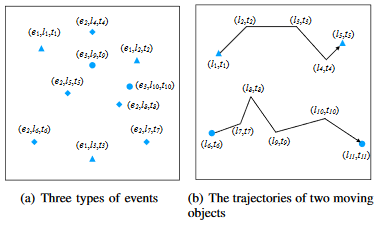}
\centering
\end{figure}
\textbf{Event Data} are events occurring at point locations and times (e.g crime reported, social media post, traffic accident, phone call, text message), and this denotes when are where the event occured. Fig. 1(a) shows an example of spatial point pattern in a two-dimensional Euclidean coordinate system, where ($l_i$, $t_i$) denotes the location and time point of an event. \par
\textbf{Trajectory Data} denotes the paths traced by objects moving in space over time (e.g bike trip, taxi trip, bus route). This is usually collected by sensors deployed on moving objects that periodically transmit the location on the moving object each time. Fig. 1(b) shows an illustration of two trajectories. Each trajectory can be characterized as a sequence [($l_1$, $t_1$), ($l_2$, $t_2$)...($l_n$, $t_n$)], where $l_i$ is the location (e.g latitude and longitude) and $t_i$ is the time when the moving object passes this location.
\subsection{Utility vs. Privacy Trade off}
This common model is considered during a data release, as the distributor, we want to ensure that there is a clear outline of what this data contains and it’s inherent importance. Such that, we only include relevant information, as well as, maintain the data’s utility for researchers. Our top priority is to keep our users identity safe while providing useful analytics. At a high level, allows engineers to evaluate possible applications for this data and consider the following questions:
\begin{enumerate}
    \item Access Control: 
    \begin{itemize}
        \item Who will have access? 
        \item How will they access the data?
    \end{itemize}
    \item Analysis:
    \begin{itemize}
        \item What is the outcome of this analysis? 
        \item What are potential auxiliary datasets?
    \end{itemize}
    \item Data: 
    \begin{itemize}
        \item Are all identifiers necessary? 
        \item What can we remove or change to limit exposure?
    \end{itemize}
\end{enumerate} \par
In general \cite{atluri_spatio-temporal_2018}, privacy is an individual concept and should be measured separately for each individual, while utility is an aggregate concept and should be measured cumulatively for all useful information. For privacy, we measure the worst case privacy loss for an individual, while utility is measured by the total usefulness of the data. With spatio-temporal event data, we can consider for each user, how revealing their location data is to recover their identity, while the total dataset reveals the number of clusters at a given point or bounded box. With trajectory data, we must acknowledge the unique movements for each user, while preserving group migration patterns. Our primary concern is the amount of new information that can be gathered about individual users from the released dataset. Any information that deviates from the truth, may lead to increased privacy, however only correct information contributes to utility. To assist our efforts in anonymizing the dataset, we must identify common assumptions about the adversary's knowledge.
\subsection{Attack Model}
% What an attacker might want to do -- reidentify a specific person, reidentify a set of people, etc.
% What an attacker might have -- external information, etc.
% What an attacker might be capable of -- direct access to the db, access to limited db queries, access to database or population statistics, etc.
% and give a citation of a case where an attack or a paper demonstrates each of these
% I'd avoid getting too technical
% so probably don't include a formal model
\begin{figure*}[ht]
  \centering
  \includegraphics[width=\textwidth,height=9cm]{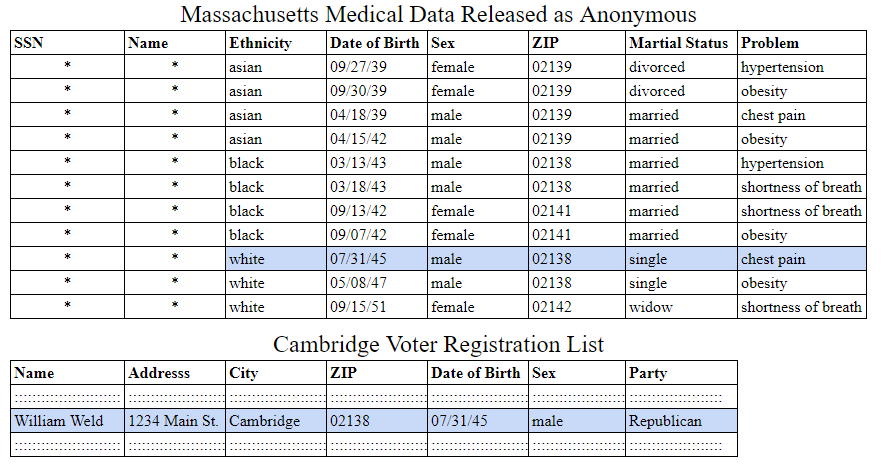}
  \caption{Adversarial use of an Auxiliary Dataset \cite{sweeney_k-anonymity_2002}}
  \label{aux}
\end{figure*}
Our attacker, Alice, wants to gather more sensitive information about a single user, such as, personal addresses, frequently visited locations, demographic, socio-economic and other attributes of the victim. This type of attack is known as the \textbf{re-identification attack} where Alice can collect enough information to identify an anonymous user from our dataset. This type of attack comes in many different forms, but re-identification is the fundamental goal of the adversary. \par
In general, a common assumption of our attacker, is that she has some \textbf{background knowledge} about her victim. Whether the attacker has some information about a single or multiple locations, they can match trajectories or event data contained inside of the dataset \cite{gramaglia_ktauepsilon-anonymity_2017}. If successful, the attacker can learn more about the unique pattern of the victim, as well as, know his/her whole trajectory, which is privacy leakage. \par
Alice could also have access to public records, social media accounts, and/or other \textbf{auxiliary datasets} to assist her efforts to create a richer dataset. As shown in Figure \ref{aux}, William Weld’s identity was exposed by limiting the number of potential patients in an “Anonymized Medical Dataset” \cite{barth-jones_re-identification_2012, sweeney_k-anonymity_2002}, due to attribute linkage of his zip code, gender, and date of birth. With raw location data posing a massive risk to user privacy, data matching is another common phenomenon with a user who uses a social media account and “checks-in” to locations upon arrival \cite{kondor_towards_2018, qian_-anonymizing_2016, wang_-anonymization_2018}. These points of interest allow an adversary to link multiple or a single user to a given location and a certain time. Another main consideration of our adversary is the \textbf{probabilistic attack} where patterns of the data repeat among various times during the day. In that, let us imagine a scenario where an adversary knows of a small set of spatio-temporal points in the trajectory of the user. When successful, it would reveal the whole trajectory of a single user, which would allow Alice to infer sensitive information such as home/work locations and daily routines. \par
A single raw point of location data can be classified from \textbf{semantic attacks}. In this, attackers can acquire an individual's private information by analyzing the meaning within the features of frequently visited locations in the dataset without attempting re-identification. Semantic features are attributes of an object that overlap such that a square is four-sided and a quadrilateral. For location data it is a point of interest that it’s attributes can infer more information about individuals such as: entertainment, education, scenery, business, industry, and residences . This type of labeling allows an adversary to target an individual based on their interest, if they are commuting, and other mobility features about our data subjects \cite{atluri_spatio-temporal_2018, tu_protecting_2019}. 

\section{Common Techniques}
There is never a blanket answer of anonymization, however there are a few techniques used to effectively limit the risk of re-identification of users. These techniques have various applications, however fall victim to simple attack models. To mitigate the many risks to anonymization, we will focus on three categories: Suppression, Generalization, and Noise. These techniques encapsulate the holistic approach during the data classification and anonymization stages, and for each category we will expand on the advantages and disadvantages of each, applicable usages and case studies, as well as, known vulnerabilities.
\subsection{Suppression}
Before we try to anonymize a given dataset, it is necessary to remove all explicit identifiers. Explicit identifiers are data types that are directly related to an individual’s identity. Examples include: Name, Social Security Number, Address, and Phone Number. There are hardly any circumstances where this data is deemed useful for an analysis. Suppression is the process of removing or replacing directly identifying information. A few common techniques include: Pseudonyms and Field Removal. These are relatively simple to implement and used frequently before a dataset can be considered to be released for research. \par
\textbf{Pseudonyms} are similar to hashing where it is a one way function to convert sensitive information to an arbitrary value. It is effective in certain situations, however vulnerable to basic attacks such as a dictionary attack. As an example, the sensitive values, phone numbers, have been set to unique user identification numbers. This relatively simple implementation obfuscates the users’ phone numbers into random or sequential numbers, and maintains the relationship between a user and their data. In this instance, a dictionary attack would not be successful in re-identifying their phone numbers, but a more sophisticated attack would target a specific location or community of users. \par
\textbf{Field removal}, in cases where the data is not necessary, replaces the value with an asterisk (empties the column). As an example, our dataset contains a timestamp, unique user id, longitude and latitude, and source and destination IP addresses. As we could pseudo anonymize the IP addresses, the information provides too much exposure for our user’s identity, due to another attribute that can be correlated with a unique user id. In this case, the most effective solution is to remove the data entirely. To emphasize the importance of suppression, Latanya Sweeney’s Attack in 1997 \cite{sweeney_k-anonymity_2002} on Governor William Weld’s medical data in Massachusetts by linking an ‘anonymized’ Medical Record with the Cambridge Voter List. \par
In the synthetic dataset shown in Fig. \ref{aux}, the governor's medical record was one of thousands of state employees within this dataset. As it was considered to be “anonymized”, the Group Insurance Commission (GIC) released this data to researchers. A researcher purchased a copy of the voter registration for Cambridge Massachusetts, and the adversary was able to re-identify a user from the voter list using his quasi-identifiers. As the Massachusetts Medical Data has successfully removed explicit identifiers such as SSN and Names, but has maintained actual birth dates. Using an auxiliary dataset, such as the Voter Registration List, we can target a specific user from these quasi-identifiers: ZIP, Gender, and Date of Birth. The adversary could take this attack further by using more than one auxiliary dataset, such as social media, a phone book, etc. Unfortunately for Governor Weld, his identity and reason for his medical visit was exposed. If the GIC generalized the date of births and zip codes, it would have made it more difficult for the adversary to re-identify individuals.
\subsection{Generalization}
Once our dataset has removed all directly identifiable information, there is still no guarantee that our users are safe. However, we can further limit the risk of re-identification through generalization. In this process, the data is made less precise, however maintains the same statistical properties of the original. The target data types are known as quasi-identifiers \cite{fung_privacy-preserving_2010} which are indirect properties of an individual. Examples include: Ethnicity, Gender, ZIP code, Marital Status, and Date of Birth. This technique is more robust to stronger attack models than suppression. As an example, k-anonymity was used to obfuscate this dataset, so that we do not disclose the users date of birth or ZIP code.
\begin{figure*}[ht]
  \includegraphics[width=\textwidth, height=4.5cm]{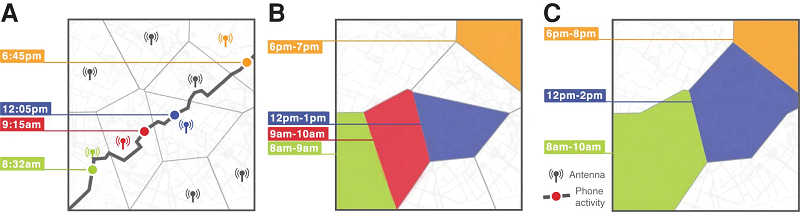}
  \caption{Obfuscation of Mobility Patterns \cite{de_montjoye_unique_2013}}
  \label{mobility}
\end{figure*}
\subsubsection{$k$-Anonymity}
The basis is to limit the possibility of re-identification by increasing the number of recurring quasi-identifiers in a given dataset. We can consider the target information in each row as a tuple of quasi-identifiers (qid), and there must be at least k repeating rows. As an example, qid = (Time, Longitude, Latitude) and k = 2, so for each tuple, it must occur at least twice in the dataset. So we can begin to cluster our users together that share this qid. As we increase our k value, the more outliers it creates in a real world dataset, such that our dataset may not have enough records to group 10 people to the same location and time to achieve 10-Anonymity. As many of our users may share similar attributes, their location and routine patterns are extremely unique, for mobility records, this may require further obfuscation of their location data points. \par
This privacy preserving technique recognizes the indistinguishability principle, where each record in a database must be identical to at least k-1 other records in the same dataset. Regarding location event data, privacy can only be achieved when each user’s trajectory or mobility patterns cannot be distinguished between k users in the dataset. A type of metric, unicity, \cite{de_montjoye_unique_2013} indicates the percentage of users who are uniquely identifiable given p points of their past trajectories. All together, the ubiquity of mobility datasets, the uniqueness of human traces and the information that can be inferred from them highlight the importance of understanding the privacy bounds of human mobility. If the attacker was to target a specific user via an identification number or other unique characteristics, they can trace their movements over time. \par
In Fig \ref{mobility}.A, the attacker could trace a single user’s call records as each dot represents the location and time the call was placed. As in Fig \ref{mobility}.B, the user’s call’s were obfuscated to the nearest cell tower and time rounded to the nearest hour. However, the attacker can draw the Voronoi lattice with respect to the cell tower reception coverage or number of calls in a cell for a given region. In Fig \ref{mobility}.C the same records rounded to the nearest two hours, as well as aggregated cell towers in batches of two to further obfuscate the user’s trajectory data. This type of k-merge technique, in ($\tau$, $\epsilon$)-Anonymity \cite{gramaglia_ktauepsilon-anonymity_2017} groups records with neighbors in close proximity such that the events at 8:32am and 9:15am are indistinguishable. The main issue with this approach is that it is an NP-Hard problem to have a global k value \cite{wong_alpha_2006} that satisfies every record in the dataset. This mainstream approach has many extensions and divergent approaches to this problem, but they maintain adherence to the indistinguishability principle of achieving k-Anonymity. \par
This model has been extensively studied because of its simplicity and effectiveness to prevent identity disclosure which occurs when an individual is linked to a particular record in the released table. However, as the dataset will not reveal who the individual is, the adversary may link new information about some individuals which is attribute disclosure. This can occur with or without identity disclosure, and usually occurs on revealing suppressed information \cite{andreou_identity_2017}, which leaves the k-anonymity model vulnerable. \par
\subsubsection{$\ell$-Diversity}
In this section, we will cover an extension on top of $k$-anonymity that addresses difficulties discussed in the previous section. To achieve $\ell$-diversity we must be aware of a lack of diversity and strong background knowledge, such that an equivalence class, a set of identical quasi-identifiers, must be at least $\ell$ “well represented” values for the sensitive attribute S. Our dataset must have sufficient diversity, for each row in the table, there must be an equal distribution of sensitive values \textit{S}. Despite our attackers' strong background knowledge, Alice needs $\ell$ - 1 harmful pieces of background information to eliminate $\ell$ - l possible sensitive values and infer a positive disclosure \cite{machanavajjhala_l-diversity_2006}. With a large variety of datasets, there are three instances of the $\ell$-diversity principle. \par
Our principle definition is known as \textbf{distinct $\ell$-diversity} where each equivalent class has at least $\ell$ well-represented sensitive values. This instance does not prevent probabilistic inference attacks \cite{altop_probabilistic_2012}, where the number of sensitive attributes are not equally distributed within the table. As an example, in one equivalence class there are ten tuples. The disease column is the sensitive attribute among this set is one “Viral Infection”, one is “Heart Disease” and the remaining eight are “Flu”. As this would satisfy 3-diversity, but the adversary can affirm that their target’s disease is “Flu” with 80\% accuracy. \par
\textbf{Entropy $\ell$-diversity} states that each equivalence class must have enough different sensitive values, but also evenly distributed within the table. Meaning that the entropy distribution for each class is at least log($\ell$). Sometimes this may be a too restrictive approach on a table with a few sensitive values in common. Therefore a large value $\ell$ implies less certainty of inferring a particular sensitive value in a group. One limitation of this approach is that it does not provide a probability based risk measurement, such that the probabilistic inference attack can still be achieved. As an example, we can classify this table into two groups, the second group {Artist, Female, [30-35)}, -¾ log(¾) - ¼ log (¼) = log(1.8). So Fig. \ref{diverse} satisfies entropy if $\ell$ $\leq$ 1.8. However, one limitation is that it does not provide a probability based risk measurement. In this case, the attacker has a 75\% probability of success to infer HIV. \par
\begin{figure}[hb]
  \includegraphics[width=8.5cm, height=3.5cm]{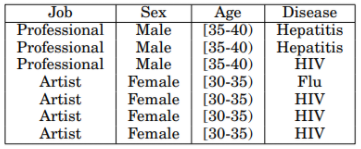}
  \caption{$\ell$-Diverse Vulnerable Dataset \cite{machanavajjhala_l-diversity_2006}}
  \label{diverse}
\end{figure}
Another instance is \textbf{recursive $\ell$-diversity} which ensures that the most frequent value does not appear too often, and the less frequent values do not appear too rarely. A qid group is (c, $\ell$) diverse if the frequency of the most frequent sensitive value is less than the sum of the number of sensitive attributes. This allows for a healthy distribution of sensitive attributes within an equivalence class. So, recursive (c, $\ell$) diversity can prevent attribute linkages, even with strong background knowledge.
\subsubsection{$t$-Closeness}
The distance between the distribution of a sensitive attribute in the equivalence class (P) and the distribution of the attribute in the whole table (Q) is no more than a threshold $t$ \cite{li_t-closeness_2007}. Requiring that $P$ and $Q$ to be close or similar would limit the amount of useful information, as well as, limit the correlation between qid attributes and sensitive attributes. If an adversary gets too clear of a picture of this correlation, then attribute disclosure occurs. The $t$ parameter in $t$-closeness enables a trade off between utility and privacy. The problem is how to measure the distance between these two probabilistic distributions which is known as variational distance. In short, we have a metric space for the attribute values so that a ground distance is defined between any pair of values. This requires Earth’s Mover Distance \cite{rubner_earth_nodate} which is based on the minimal cost that must be paid to transform one distribution into another. This method remedies the limitations of $\ell$-diversity by requiring the sensitive attribute distribution in each equivalence class to close to the overall dataset.\par
The similarity attack is when sensitive attributes in an equivalence class are distinct but have similar meaning, and our adversary, Alice, can learn important information. This attack is prevented, for Alice cannot infer that Bob has a low salary or Bob has stomach related diseases. While $t$-closeness protects against attribute disclosure, it does not prevent identity disclosure. Even with such a generalized table, we still maintain the relationship between quasi-identifiers and their corresponding sensitive attributes. Thus, it may be desirable to use both $t$-closeness and $k$-anonymity. When combined, $t$-closeness protects against homogeneity and background knowledge attacks. The drawbacks to t-closeness are identity linkage and probabilistic attacks, as well as, limited applications. 
\subsection{Noise}
Perturbation distorts the data by adding noise, aggregating values, swapping values, or generating synthesized data based on statistical properties of the original \cite{mivule_utilizing_2013}. Since these records do not correspond to real-world owners, the attacker cannot perform the same sensitive linkage attacks found with generalization. However, too much noise can make a dataset useless for analysis. \par
\textbf{Data Swapping} is an intuitive approach involving transformations that map the original data matrix into a new database which exhibits the same statistics \cite{moore_controlled_nodate, reiss_practical_nodate}. Some advantages of swapping are first, it removes the relationship between the user and their micro data. Second, it can be used on sensitive values without modifying the quasi-identifiers. Third, it is relatively simple to implement. Fourth, swapping less occurring (outliers) sensitive values provides more protection, because more common sensitive values are less likely to be of value to an adversary. Finally, it is not limited to any classification of data types. As an example, in Fig. \ref{swap} the original dataset was swapped with two different data types to obfuscate the user from their data. However, the statistical properties in both of the altered datasets have been ruined. Take the Race Swap, as the race-occupation hasn't affected the statically properties, the distribution of occupation to salary have been distorted, such that, our white clerk makes more than other executives, which is wrong. In implementation, this method chooses outliers for the data swap to make them more "normal", however can ruin the integrity of the data and render it useless for further analysis.
\begin{figure}[hb]
    \includegraphics[width=8.85cm,height=3.75cm]{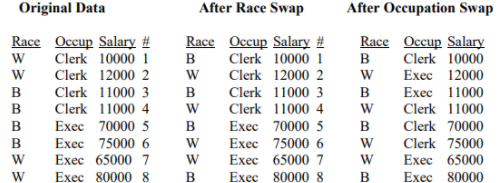}
    \caption{Consequences of Data Swapping \cite{moore_controlled_nodate}}
    \label{swap}
\end{figure}
Another technique commonly used in data distribution is through additional noise for each query to a remote database. This ensures that the data received by the third party does not directly match the data stored internally. This transformation of sensitive values works by adding or multiplying a stochastic or randomized value. The stochastic value is chosen from the "normal" distribution with zero mean and a diminutive standard deviation \cite{fung_privacy-preserving_2010, mivule_utilizing_2013}. \textbf{Additive noise}, also known as white noise, is adding a stochastic value to the original sensitive value which then replaces the original for publication, and this is known as the the transformed data point, which preservers the mean and co-variance \cite{domingo-ferrer_security_2004} of the uniform distributed noise. Another addition of noise is through \textbf{multiplicative noise} where each data point is multiplied by a chosen stochastic value within short Gaussian distribution \cite{kim_multiplicative_nodate,dwork_differential_2011}, however it is restricted during the generation of random values that must have a mean = 1 and little variance. The stochastic value ($\epsilon$) is randomly generated as a set and the mean($\mu$) and variance($\sigma^2$) can be calculated. An important note is that standard deviation is the square root of variance, so $\sigma$ is standard deviation and $\sigma^2$ is variance. \par
\subsubsection{Stochastic Noise}
\begin{figure}[th]
    \includegraphics[width=0.45\textwidth]{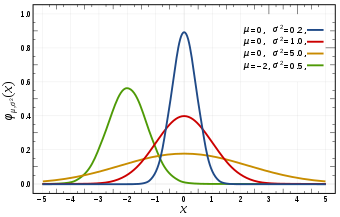}
    \caption{Various Gaussian Distributions}
    \label{guassian}
\end{figure}
In Fig. \ref{guassian}, the graph shows a variety of normal, Gaussian, distributions, with a mean($\mu$) = [-1,0] and standard deviation($\sigma$) = [0.5,2] inclusively. As the standard deviation decreases with a mean of $x$, the probability that the average of random values being $x$ approaches 1. To apply this, consider that our mean is our target noise value, and the standard deviation is our loss function. In that when using the additive noise mechanism, it maintains the mean and variance of the original data point for all transformed data points and can be thought of as an offset. With multiplicative noise, the stochastic value must be adjusted for every data point to have a mean = 1 within this short Gaussian distribution with minimal variance. \par
With a relatively intuitive design, privacy can be measured by the variance between the original and distributed data points, as well as, the noise variance for each query from a remote database. However, similar sensitive values, depending on the mechanism, will be distributed with the same variance, can be recovered from the randomized data when correlation among attributes is high but the noise is not \cite{kargupta_privacy_2003}. In that, an adversary can attempt to calculate the variance and implementation by conducting multiple queries to the remote database. 
\subsubsection{Differential Privacy}
\begin{figure}[h]
    \centering
    \includegraphics[width=8.8cm, height=4.5cm]{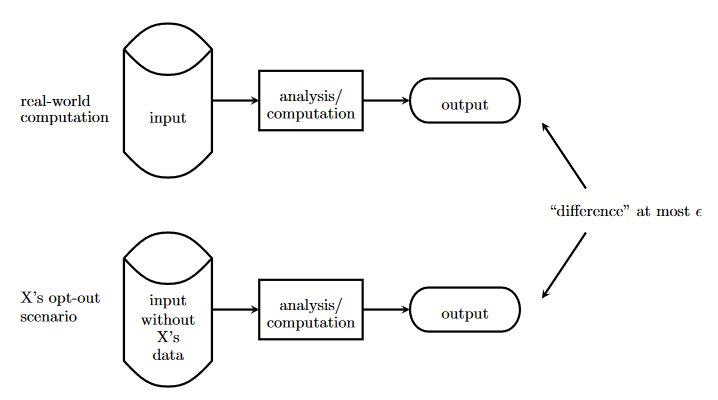}
    \caption{The maximum distance $\epsilon$ between the opt-out scenario and real-world computation should for each individual $X$ whose information is included in the input.}
    \label{diffprivacy}
\end{figure}
In the sense of statistical and machine learning analysis \cite{wood_differential_2018}, differential privacy is a strong, mathematical concept of privacy. It is used to allow a wide range of statistical estimates, such as averages, contingency tables and synthetic data, to be obtained, analyzed and exchanged based on personally identifiable information while maintaining the privacy of the individuals in the data. The guarantee that anyone seeing the result of a differentially private analysis will ultimately make the same inferences about any given individual’s private information whether or not that individual’s private information is included in the input to the analysis. The privacy loss parameter ($\epsilon$) limits what can be learned about an individual as a result of the private information being included in the analysis. This privacy loss can grow as information is used in multiple analyses, but is bounded as a function of $\epsilon$ and number of analyses performed. Thus differential privacy guarantee can be understood as:
\begin{itemize}
    \item Protects an individual’s information as if it was not used in the analysis at all.
    \item Ensures that an individual’s data will not reveal any PII that is specific to the data subject, unless the individual’s information was used in the analysis. 
    \item Masks the contribution of any single individual, making it impossible to infer any specific information about an individual.  
\end{itemize}
Differential Privacy addresses the paradox of learning nothing about an individual while learning useful information about a population \cite{dwork_differential_2008, dwork_algorithmic_2014}. Informally, this mechanism ensures privacy, in that, the output of a differentially private analysis will be roughly the same whether an individual contributes their data or not as you can see in Fig. \ref{diffprivacy}. 
A high-level definition, by Cynthia Dwork, consists of a randomization response which captures some property $P$ of an individual. As an example, in a social study of binary input of positive (Yes) or negative (No) responses, individuals who report their answer undergo the following process:
\begin{enumerate}
    \item Flip a coin
    \item If tails, then respond truthfully
    \item If heads, then flip again
    \begin{enumerate}
        \item If heads, respond Yes
        \item If tails, respond No
    \end{enumerate}
\end{enumerate}
The privacy assurance stems from plausible deniability where there is a 1/4 probability that an individual actually has a property $P$. The “random” mechanism in this case is with the probability of the resulting face of flipping a coin twice. \\
Heads on the first toss, and tails on the second toss:
\begin{equation}
\text{} \\
\begin{aligned}
= P(Yes \mid No) \\
= P(\text{tails on first}) * P(\text{heads on second}) \\
= 0.5*0.5 = 0.25
\end{aligned}
\end{equation}
Known first toss was heads and heads on the second toss:
\begin{equation}
\begin{aligned}
P(Yes \mid Yes) \\
= P(\text{heads on first}) + \\
P(\text{heads on first})*P(\text{heads on second}) \\
= 0.5 + 0.5*0.5 = 0.75
\end{aligned}
\end{equation}
Accuracy comes from the noise generation procedure where the expected ratio of positive responses are a 1/4 of individuals who do not have this property plus 3/4 of individuals having the property $P$. Thus, our truth fraction of responses $t$ are equivalent to:
\begin{equation}
(1/4)(1-t) + (3/4)t = (1/4) + t/2
\end{equation}
and due to our data analysis only considers a binary set of responses from individuals, we can estimate $t$ as twice the number of responses subtracted by 1/2 :
\begin{equation}
2((1/4) + t/2) - 1/2
\end{equation}
A more formal definition of Differential Privacy consists of the following components:
\begin{itemize}
    \item Randomization mechanism that queries information from a database and incorporates some noise.
    \item Internal parameter of all potential output of the randomization mechanism that could be predicted.
    \item Two parallel databases (x, y) where one dataset (x) has N records and the other (y) has N-1 records.
    \item $\epsilon$, the maximum distance between a query on database (x) and the same query on database (y).
    \item $\delta$, the probability of information accidentally being leaked.
\end{itemize}
The most important components are it’s parameters, for $\epsilon$ is our privacy parameter/budget, and $\delta$ is our probability of exposing identities within the dataset. In general, the smaller our privacy budget is while exposure is zero, the output is differentially private. However as $\epsilon$ is the maximum distance between the same query on each dataset, the smaller it is, the more similar the outputs will be, therefore, proves higher level of privacy when compared to a larger $\epsilon$ privacy budget. The probability of information accidentally being leaked ($\delta$) that is a constant, so it dictates a threshold of at least 1-$\delta$ probability of privacy loss. \par 
($\epsilon$, 0)-differential privacy ensures, for every run of the randomization mechanism, the output on the first dataset (x) is almost equal to the output in the neighboring dataset (y) \cite{dwork_algorithmic_2014}. Thus, the probability of data leaks is zero, and a differentially private set with a sensitivity of less than 1 will have very similar output. \par
($\epsilon$, $\delta$)-differential privacy ensures for every pair of neighboring dataset the absolute value of the privacy loss with be bounded by $\epsilon$ with probability at least 1-$\delta$ \cite{mivule_utilizing_2013}. In this case, the chances of data leaks are possible, due to a higher sensitivity ($\epsilon$) or large database containing many attributes. Thus, if the sensitivity is not small, the less likely the outputs of the neighboring datasets will be the same. \par
\begin{figure}[th]
    \centering
    \includegraphics[width=0.4\textwidth]{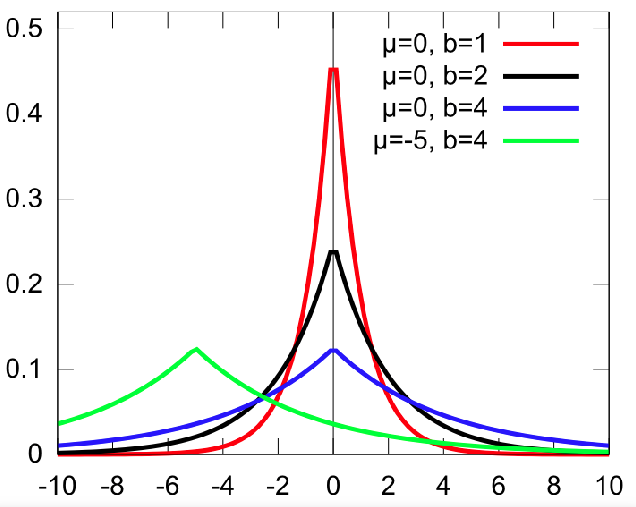}
    \caption{Various Laplace Distributions}
        \label{laplace}
\end{figure}
The Laplace distribution function looks like the exponential distribution with a copy reflected across the y-axis. In Fig. \ref{laplace} $\mu$ represents the shift of the probability distribution, and $b$ is a scalar parameter. This mechanism is effective on small, normally distributed datasets where there are minimal outliers, and it can easily mask a single contribution (one record difference in x and y). However, on a dataset with many fluctuations and outliers, adds way too much noise. Such that, it becomes exponentially more challenging to achieve Laplace distributions \cite{near_differential_2020, wood_differential_2018}. \par
The differential privacy technique used by Apple is based on the idea that statistical noise that is slightly biased can mask a user's identity. Such that, if many people submit the same data, the noise that has been added will average out over a large set of data points. They utilize a count mean sketch to determine the most popular emojis from the user's keyboard and typing history. For the user, it appears that their favorite emoji is readily available on their keyboard, but on the back end, the count mean sketch uses variations of SHA-256 hash functions. Once the data is encoded with the hash function, each coordinate of the vector (representation of the keyboard) is then flipped with a probability of 1/(1+$e^{\epsilon/2}$), where $\epsilon$ is the privacy parameter. To stay within the privacy budget, a strict budget of contributions per user, a random vector of this sketch matrix is sent to Apple. Furthermore, they also utilize a Hadamard Count Mean Sketch which extends the previous method with noise injection. It applies a type of mathematical operation called the Hadamard basis transformation, in which it samples 1 bit at random to send rather than an entire row in the Count Mean Sketch technique. This reduces communication cost to 1 bit at the expense of accuracy.
\section{Data Release Policies}
In the past decade, multiple telecommunication companies have explored ways to share their datasets with researchers. In this section we review these data sharing efforts categorized in  three groups by~\cite{de2018privacy} and describe the limitation of each category.

\subsection{Limited Access:} This group of work refers to the cases where remote access to the data is provided to researchers often referred to as the privacy-through-security approach. The data are not released but instead stay within the premises and under the control of the operator. This model requires signed NDA and DUA (data use agreement) with the controller. The controller can supervise who accesses the data how the data are being used. Such a supervision includes active monitoring of the secured environment and/or controlling the output. This strategy can ensure that no individual-level or raw data leave the server. While they do not remove all possible risks, these security-based mechanisms already strongly limit the risks of the data to be re-identified and misused.  This model was used by Flowminder\footnote{https://web.flowminder.org} where a small number of registered researchers analyzed pseudo-anonymized mobile phone data remotely with security measures in place to study people’s mobility directly after the Nepal earthquake. A main challenge with this model is that it requires a significant human investment from the telecom company.

\subsection{Limited Release:} This model refers to a category of data sharing where the mobile phone data is transformed in house and is given to third parties under a legal contract. Transformation process aims to address re-identification concerns and has been historically achieved through the following:
 Masking the exact location of the cell towers and antennas, by randomly shifting them by a few meters. 
Data sampling: sub sampling users id and longitudinal re-sampling with new identifiers.
Coarsening the spatial data to correspond to the Voronoi~\cite{de20198} translation of the antenna as opposed to the exact latitude and longitude. 

An example of limited releases are Orange’s D4D challenges~\cite{blondel2012data} and Turk Telecom D4R~\cite{salah2018data}, in which data were sample and the spatial granularity was coarsened before being released to selected teams of researchers under strict DUA. 
In general any transformations affect the quality and quantity of the data and therefore the statistical inferences that can be made and potentially preventing important research questions to be explored.  Spatial and temporal coarsening however has been shown to only marginally help prevent re-identification while restricting the general use of the data~\cite{de2013unique}. The main implementation challenge of the limited release model is probably the choice of the transformation. It requires an in-depth understanding of all of the current but also future uses of the data, as anonymization can usually only be performed once.

\subsection{Pre-computed indicators and synthetic data:}
In this model a set of pre-computed indicators are derived from mobile phone data or synthetic data and  released to third-parties. These indicators could be computed in    individual level, corresponding to the number of calls, or alternatively aggregated across individuals to capture for example the number of users per tower over time, long or short-term mobility matrices, and matrices of inter-towers communications.

As pre-computed indicators are often aggregated, in general they are more disconnected from both the raw and potential auxiliary data. This property provides strong anonymization although there exists recent debates as to whether such aggregation methods are sufficient to provide privacy guarantees as they fail to address “group privacy”~\cite{bloustein2018individual}.  Pre-computed indicators were previously used by Orange Telecom in D4D challenge for the open release of well-established and stable-across-time mobility metric.  Other examples include the release of flow maps parameterized using mobile phone data by Flowminder as part of the fight against Ebola~\cite{wesolowski2014commentary} and the release of tourism statistics by Statistics Netherlands~\cite{heerschap2014innovation}. 
Similarly, synthetic data representations can be parameterized using mobile phone data. Such that it maintains the statistical properties of the original data without compromising individual’s privacy.   However, little work, so far, exists in synthetic mobile phone data representations.

% Design/Methodology
\section{Recommendations}
To date, there are no “one size fits all” approaches to privacy; however, there are guidelines that can be used to limit risk when releasing data to researchers.  It is our recommendation that data releases follow a defense-in-depth approach to privacy that leverages both technical and non-technical approaches to privacy.
\subsection{Non-Technical}
The first line of defense against a privacy breach is to understand the intended use of your data.  It has been noted in this paper that data privacy may be lost even when the current state of the art privacy techniques are applied.  It would be naive to believe that all attacks against privacy have been discovered at this point.  Furthermore, depending on the resources available, an adversary may have enough external information to join with a released dataset to identify unique individuals in the dataset.  To address this, non-technical approaches can provide risk-limiting solutions. \par
These datasets should not be made freely available without some barrier to acquire them.  We recommend that researchers must identify the research question(s) they will answer by using the data, the data they will need for their research, and how they will use that data to achieve their goals.  In doing this, the cost to a potential attacker will be raised in terms of effort and self-identification.  Furthermore, it allows the Carrier to filter out researchers who are not trying to be malicious, but whose work intends to violate the privacy of the users in the datasets.  As a second non-technical defense to strengthen the first, researchers should sign an agreement that they will not attempt to deanonymize any of the data in the dataset.
\subsection{Technical}
This paper has presented a variety of technical approaches to securing datasets.  Many of these techniques are well-known while others are still in the process of being adopted.  One of the issues with applying these techniques is that it requires an understanding of which techniques are most appropriate for which situations.  Many of these techniques balance trade-offs between utility and privacy that require trial and error to decide which parameters to configure them with. \par
Our recommendations is that a internal, data-specific tool is developed that can assist the Carrier in deciding the most appropriate privacy-preserving technologies to apply to a given dataset for a specific application (as proposed by the researchers in the non-technical recommendation section).  Such a tool could use AI-based approaches to analyze both the needs of the researcher and the data included in the dataset to suggest which technique is most appropriate.  Furthermore, it would be extensible so that new techniques could be developed and added over time as they are developed by the research community.
% References 
\bibliography{references.bib}{}

\begin{thebibliography}{10}

\bibitem{noauthor_california_2018}
California {Consumer} {Privacy} {Act} ({CCPA}), October 2018.

\bibitem{altop_probabilistic_2012}
Baris Altop, Mehmet~Ercan Nergiz, and Yücel Saygin.
\newblock A {Probabilistic} {Inference} {Attack} on {Suppressed} {Social}
  {Networks}.
\newblock In {\em 2012 {IEEE}/{ACM} {International} {Conference} on {Advances}
  in {Social} {Networks} {Analysis} and {Mining}}, pages 726--727, August 2012.

\bibitem{andreou_identity_2017}
Athanasios Andreou, Oana Goga, and Patrick Loiseau.
\newblock Identity vs. {Attribute} {Disclosure} {Risks} for {Users} with
  {Multiple} {Social} {Profiles}.
\newblock In {\em Proceedings of the 2017 {IEEE}/{ACM} {International}
  {Conference} on {Advances} in {Social} {Networks} {Analysis} and {Mining}
  2017}, pages 163--170, Sydney Australia, July 2017. ACM.

\bibitem{atluri_spatio-temporal_2018}
Gowtham Atluri, Anuj Karpatne, and Vipin Kumar.
\newblock Spatio-{Temporal} {Data} {Mining}: {A} {Survey} of {Problems} and
  {Methods}.
\newblock {\em ACM Computing Surveys}, 51(4):1--41, September 2018.
\newblock Number: 4.

\bibitem{barth-jones_re-identification_2012}
Daniel~C. Barth-Jones.
\newblock The '{Re}-{Identification}' of {Governor} {William} {Weld}'s
  {Medical} {Information}: {A} {Critical} {Re}-{Examination} of {Health} {Data}
  {Identification} {Risks} and {Privacy} {Protections}, {Then} and {Now}.
\newblock {\em SSRN Electronic Journal}, 2012.

\bibitem{blondel2012data}
Vincent~D Blondel, Markus Esch, Connie Chan, Fabrice Cl{\'e}rot, Pierre
  Deville, Etienne Huens, Fr{\'e}d{\'e}ric Morlot, Zbigniew Smoreda, and Cezary
  Ziemlicki.
\newblock Data for development: the d4d challenge on mobile phone data.
\newblock {\em arXiv preprint arXiv:1210.0137}, 2012.

\bibitem{bloustein2018individual}
Edward~J Bloustein and Nathaniel~J Pallone.
\newblock {\em Individual and group privacy}.
\newblock Routledge, 2018.

\bibitem{de2018privacy}
Yves-Alexandre de~Montjoye, S{\'e}bastien Gambs, Vincent Blondel, Geoffrey
  Canright, Nicolas De~Cordes, S{\'e}bastien Deletaille, Kenth Eng{\o}-Monsen,
  Manuel Garcia-Herranz, Jake Kendall, Cameron Kerry, et~al.
\newblock On the privacy-conscientious use of mobile phone data.
\newblock {\em Scientific data}, 5(1):1--6, 2018.

\bibitem{de2013unique}
Yves-Alexandre De~Montjoye, C{\'e}sar~A Hidalgo, Michel Verleysen, and
  Vincent~D Blondel.
\newblock Unique in the crowd: The privacy bounds of human mobility.
\newblock {\em Scientific reports}, 3:1376, 2013.

\bibitem{de_montjoye_unique_2013}
Yves-Alexandre de~Montjoye, César~A. Hidalgo, Michel Verleysen, and Vincent~D.
  Blondel.
\newblock Unique in the {Crowd}: {The} privacy bounds of human mobility.
\newblock {\em Scientific Reports}, 3(1):1--5, March 2013.

\bibitem{de20198}
Yves-Alexandre de~Montjoye, Jake Kendall, and Cameron~F Kerry.
\newblock 8 enabling humanitarian use of mobile phone data.
\newblock {\em Trusted Data: A New Framework for Identity and Data Sharing},
  page 167, 2019.

\bibitem{domingo-ferrer_security_2004}
Josep Domingo-ferrer, Francesc Sebe, and Jordi Castella-Roca.
\newblock On the {Security} of {Noise} {Addition} for {Privacy} in
  {Statistical} {Databases}.
\newblock In {\em Privacy in {Statistical} {Databases} 2004}, pages 149--161.
  Springer, 2004.

\bibitem{dwork_differential_2008}
Cynthia Dwork.
\newblock Differential privacy: a survey of results.
\newblock In {\em Proceedings of the 5th international conference on {Theory}
  and applications of models of computation}, {TAMC}'08, pages 1--19, Berlin,
  Heidelberg, April 2008. Springer-Verlag.

\bibitem{dwork_differential_2011}
Cynthia Dwork.
\newblock Differential {Privacy}.
\newblock In Henk C.~A. van Tilborg and Sushil Jajodia, editors, {\em
  Encyclopedia of {Cryptography} and {Security}}, pages 338--340. Springer US,
  Boston, MA, 2011.

\bibitem{dwork_algorithmic_2014}
Cynthia Dwork and Aaron Roth.
\newblock The {Algorithmic} {Foundations} of {Differential} {Privacy}.
\newblock {\em Foundations and Trends® in Theoretical Computer Science},
  9(3–4):211--407, August 2014.
\newblock Publisher: Now Publishers, Inc.

\bibitem{fung_privacy-preserving_2010}
Benjamin C.~M. Fung, Ke~Wang, Rui Chen, and Philip~S. Yu.
\newblock Privacy-preserving data publishing: {A} survey of recent
  developments.
\newblock {\em ACM Computing Surveys}, 42(4):1--53, June 2010.

\bibitem{gramaglia_ktauepsilon-anonymity_2017}
Marco Gramaglia, Marco Fiore, Alberto Tarable, and Albert Banchs.
\newblock
  \$k{\textasciicircum}\{{\textbackslash}tau,{\textbackslash}epsilon\}\$-anonymity:
  {Towards} {Privacy}-{Preserving} {Publishing} of {Spatiotemporal}
  {Trajectory} {Data}.
\newblock {\em arXiv:1701.02243 [cs]}, January 2017.
\newblock arXiv: 1701.02243.

\bibitem{heerschap2014innovation}
Nico Heerschap, Shirley Ortega, Alex Priem, and May Offermans.
\newblock Innovation of tourism statistics through the use of new big data
  sources.
\newblock In {\em 12th global forum on tourism statistics, Prague, CZ}, volume
  716, 2014.

\bibitem{hong_towards_2018}
Lingzi Hong, Myeong Lee, Afra Mashhadi, and Vanessa Frias-Martinez.
\newblock Towards {Understanding} {Communication} {Behavior} {Changes} {During}
  {Floods} {Using} {Cell} {Phone} {Data}.
\newblock In {\em Social {Informatics}}, pages 97--107. Springer, Cham,
  September 2018.

\bibitem{kargupta_privacy_2003}
H.~Kargupta, S.~Datta, Q.~Wang, and {Krishnamoorthy Sivakumar}.
\newblock On the privacy preserving properties of random data perturbation
  techniques.
\newblock In {\em Third {IEEE} {International} {Conference} on {Data}
  {Mining}}, pages 99--106, November 2003.

\bibitem{kim_multiplicative_nodate}
Jay~J Kim and William~E Winkler.
\newblock Multiplicative {Noise} for {Masking} {Continuous} {Data}.
\newblock page~18.

\bibitem{kondor_towards_2018}
D.~Kondor, B.~Hashemian, Y.~de Montjoye, and C.~Ratti.
\newblock Towards matching user mobility traces in large-scale datasets.
\newblock {\em IEEE Transactions on Big Data}, pages 1--1, 2018.

\bibitem{li_t-closeness_2007}
Ninghui Li, Tiancheng Li, and Suresh Venkatasubramanian.
\newblock t-{Closeness}: {Privacy} {Beyond} k-{Anonymity} and l-{Diversity}.
\newblock In {\em 2007 {IEEE} 23rd {International} {Conference} on {Data}
  {Engineering}}, pages 106--115, April 2007.
\newblock ISSN: 2375-026X.

\bibitem{machanavajjhala_l-diversity_2006}
A.~Machanavajjhala, J.~Gehrke, D.~Kifer, and M.~Venkitasubramaniam.
\newblock L-diversity: privacy beyond k-anonymity.
\newblock In {\em 22nd {International} {Conference} on {Data} {Engineering}
  ({ICDE}'06)}, pages 24--24, April 2006.
\newblock ISSN: 2375-026X.

\bibitem{mivule_utilizing_2013}
Kato Mivule.
\newblock Utilizing {Noise} {Addition} for {Data} {Privacy}, an {Overview}.
\newblock {\em arXiv:1309.3958 [cs]}, September 2013.
\newblock arXiv: 1309.3958.

\bibitem{moore_controlled_nodate}
Richard~A Moore.
\newblock {CONTROLLED} {DATA}-{SWAPPING} {TECHNIQUES} {FOR} {MASKING} {PUBLIC}
  {USE} {MICRODATA} {SETS}.
\newblock page~42.

\bibitem{near_differential_2020}
Joseph Near, David Darais, and Kaitlin Boeckl.
\newblock Differential {Privacy} for {Privacy}-{Preserving} {Data} {Analysis}:
  {An} {Introduction} to our {Blog} {Series}, July 2020.
\newblock Last Modified: 2020-08-04T14:13-04:00.

\bibitem{european_parliament_and_council_of_european_union_art_2016}
European Parliament {and}~Council of~European~Union.
\newblock Art. 4 - {Definitions}, 2016.

\bibitem{oliver_mobile_2020}
Nuria Oliver, Bruno Lepri, Harald Sterly, Renaud Lambiotte, Sébastien
  Deletaille, Marco De~Nadai, Emmanuel Letouzé, Albert~Ali Salah, Richard
  Benjamins, Ciro Cattuto, Vittoria Colizza, Nicolas de~Cordes, Samuel~P.
  Fraiberger, Till Koebe, Sune Lehmann, Juan Murillo, Alex Pentland, Phuong~N
  Pham, Frédéric Pivetta, Jari Saramäki, Samuel~V. Scarpino, Michele
  Tizzoni, Stefaan Verhulst, and Patrick Vinck.
\newblock Mobile phone data for informing public health actions across the
  {COVID}-19 pandemic life cycle.
\newblock {\em Science Advances}, 6(23):eabc0764, June 2020.
\newblock Number: 23.

\bibitem{oliver_mobile_2015}
Nuria Oliver, Aleksandar Matic, and Enrique Frias-Martinez.
\newblock Mobile {Network} {Data} for {Public} {Health}: {Opportunities} and
  {Challenges}.
\newblock {\em Frontiers in Public Health}, 3, August 2015.

\bibitem{qian_-anonymizing_2016}
J.~Qian, X.~Li, C.~Zhang, and L.~Chen.
\newblock De-anonymizing social networks and inferring private attributes using
  knowledge graphs.
\newblock In {\em {IEEE} {INFOCOM} 2016 - {The} 35th {Annual} {IEEE}
  {International} {Conference} on {Computer} {Communications}}, pages 1--9,
  April 2016.

\bibitem{reiss_practical_nodate}
Steven~P Reiss.
\newblock Practical {Data}-{Swapping}: {The} {First} {Steps}.
\newblock {\em ACM Transactions on Database Systems}, 9(1):18.
\newblock Number: 1.

\bibitem{rubner_earth_nodate}
Yossi Rubner, Carlo Tomasi, and Leonidas~J Guibas.
\newblock The {Earth} {Mover}'s {Distance} as a {Metric} for {Image}
  {Retrieval}.
\newblock page~20.

\bibitem{salah2018data}
Albert~Ali Salah, Alex Pentland, Bruno Lepri, Emmanuel Letouz{\'e}, Patrick
  Vinck, Yves-Alexandre de~Montjoye, Xiaowen Dong, and Ozge Dagdelen.
\newblock Data for refugees: the d4r challenge on mobility of syrian refugees
  in turkey.
\newblock {\em arXiv preprint arXiv:1807.00523}, 2018.

\bibitem{smith_ubiquitous_nodate}
Christopher Smith, Afra Mashhadi, and Licia Capra.
\newblock {\em Ubiquitous {Sensing} for {Mapping} {Poverty} in {Developing}
  {Countries}}.

\bibitem{sweeney_k-anonymity_2002}
Latanya Sweeney.
\newblock k-{ANONYMITY}: {A} {MODEL} {FOR} {PROTECTING} {PRIVACY}.
\newblock {\em International Journal of Uncertainty, Fuzziness and
  Knowledge-Based Systems}, 10(05):557--570, October 2002.
\newblock Number: 05.

\bibitem{tu_protecting_2019}
Z.~Tu, K.~Zhao, F.~Xu, Y.~Li, L.~Su, and D.~Jin.
\newblock Protecting {Trajectory} {From} {Semantic} {Attack} {Considering}
  \$k\$ -{Anonymity}, \$l\$ -{Diversity}, and \$t\$ -{Closeness}.
\newblock {\em IEEE Transactions on Network and Service Management},
  16(1):264--278, March 2019.
\newblock Conference Name: IEEE Transactions on Network and Service Management.

\bibitem{wang_-anonymization_2018}
Huandong Wang.
\newblock De-anonymization algorithm of mobility trajectories:
  whd14/{De}-anonymization-of-{Mobility}-{Trajectories}, December 2018.
\newblock original-date: 2018-03-02T08:48:01Z.

\bibitem{wang_deep_2019}
Senzhang Wang, Jiannong Cao, and Philip~S. Yu.
\newblock Deep {Learning} for {Spatio}-{Temporal} {Data} {Mining}: {A}
  {Survey}.
\newblock {\em arXiv:1906.04928 [cs, stat]}, June 2019.
\newblock arXiv: 1906.04928.

\bibitem{wesolowski2014commentary}
Amy Wesolowski, Caroline~O Buckee, Linus Bengtsson, Erik Wetter, Xin Lu, and
  Andrew~J Tatem.
\newblock Commentary: containing the ebola outbreak-the potential and challenge
  of mobile network data.
\newblock {\em PLoS currents}, 6, 2014.

\bibitem{wong_alpha_2006}
Raymond Chi-Wing Wong, Jiuyong Li, Ada Wai-Chee Fu, and Ke~Wang.
\newblock (\${\textbackslash}alpha\$, {K})-anonymity: {An} {Enhanced}
  {K}-anonymity {Model} for {Privacy} {Preserving} {Data} {Publishing}.
\newblock In {\em Proceedings of the 12th {ACM} {SIGKDD} {International}
  {Conference} on {Knowledge} {Discovery} and {Data} {Mining}}, {KDD} '06,
  pages 754--759, New York, NY, USA, 2006. ACM.
\newblock event-place: Philadelphia, PA, USA.

\bibitem{wood_differential_2018}
Alexandra Wood, Micah Altman, Aaron Bembenek, Mark Bun, Marco Gaboardi, James
  Honaker, Kobbi Nissim, David O'Brien, Thomas Steinke, and Salil Vadhan.
\newblock Differential {Privacy}: {A} {Primer} for a {Non}-{Technical}
  {Audience}.
\newblock {\em SSRN Electronic Journal}, 2018.

\end{thebibliography}
\bibliographystyle{plain}
\end{document}